\documentclass[preprint,chaos,amsmath,amsthm,amssymb,showpacs,floatfix]{revtex4-1}
\usepackage{graphicx}% Include figure files
\usepackage{dcolumn}% Align table columns on decimal point
\usepackage{bm}% bold math
\newcommand{\be}{\begin{equation}}
\newcommand{\ee}{\end{equation}}

\newtheorem{theorem}{Theorem}%[section]

\begin{document}
\title{ Caputo Standard $\alpha$-Family of Maps: 
Fractional Difference vs. Fractional}

\author{M. Edelman}
%\email{edelman@cims.nyu.edu}

\affiliation{Department of Physics, Stern College at Yeshiva University, 245
  Lexington Ave, New York, NY 10016, USA 
\\ Courant Institute of
Mathematical Sciences, New York University, 251 Mercer St., New York, NY
10012, USA
}

\date{\today}

\begin{abstract}
In this paper the author compares behaviors of systems 
which can be described by fractional differential and 
fractional difference  equations using the fractional 
and fractional difference Caputo Standard $\alpha$-Families 
of Maps as examples. 
The author shows that properties of fractional difference maps 
(systems with falling factorial-law memory) are similar to the 
properties of fractional maps (systems with power-law memory). 
The similarities (types of attractors, power-law convergence 
of trajectories, existence of cascade of bifurcations and 
intermittent cascade of bifurcations type trajectories, and 
dependence of properties on the memory parameter $\alpha$) 
and differences in properties of falling factorial- and 
power-law memory maps are investigated.  
\end{abstract}
\maketitle

%discrete map \sep fractional difference \sep attractor 

%\PACS 05.45.Pq \sep  45.10.Hj  
%\end{keyword}
%\end{frontmatter}

{\bf Unlike fractional calculus, whose history is more than three hundred 
years old, fractional difference calculus is relatively young - it is 
approximately thirty years old. This is probably the result of the fact 
that, despite the beautiful mathematics which arises during the
development of fractional difference calculus, it doesn't have too many 
applications in nature and engineering. As it has been recently
demonstrated, the simplest fractional difference equations 
(when a fractional difference on the left is equal to a nonlinear 
function on the right) are equivalent to maps with falling factorial-law 
memory. Falling factorial-law memory is asymptotically power-law memory 
with the rate of convergence proportional to the inverse of time (or
number of iterations in discrete cases). It is difficult to distinguish 
power-law from asymptotically power-law memory which frequently appears in
investigation of noisy natural systems. This is the major motivation for 
the presented work in which we study the simplest fractional difference 
equations with sine nonlinearity and compare their properties with 
properties of the corresponding systems with power-law memory.}

\section{Introduction}

Systems with memory are common in biology, 
social sciences, physics, and engineering (see review \cite{MyReview}). 
Systems with power-law memory in many cases can be described by fractional
differential equations \cite{SKM,KST,Podlubny}. If a natural system is a
discrete one and can be described by a fractional difference equation, then
the system's memory is falling factorial-law memory \cite{GZ,MR,Atici}, which
is asymptotically power-law memory \cite{IFAC}.

To study nonlinear systems with power-law memory Tarasov and 
Zaslavsky \cite{TZ} introduced fractional maps, which are equivalent 
to the fractional differential equations of nonlinear systems experiencing
periodic delta function-kicks. Fractional Riemann-Liouville and Caputo 
Standard Maps corresponding to the fractional differential equations 
with  orders of derivatives $\alpha>1$ were used to 
investigate general properties of 
fractional dynamical systems in \cite{TZ,FM1,FM2,FM3,FM4,FM5,FM6}. 
The notion of fractional $\alpha$-families of maps ($\alpha$FM), which 
allowed the study of fractional Standard and Logistic Maps corresponding 
to $\alpha>0$, was introduced later in \cite{MyReview,MEDNC,MEChaos}.

Fractional difference equations were investigated in many papers (see, e.g., 
\cite{GZ,MR,Atici,Aga,Anastas,DifSum,Fall,FallC}).
The authors of \cite{IFAC,DifSum,Fall,FallC}
demonstrated that in some cases fractional difference equations are equivalent 
to maps  with falling factorial-law memory (which we will call fractional
difference maps), where 
falling factorial function is defined as 
\begin{equation}
t^{(\alpha)} =\frac{\Gamma(t+1)}{\Gamma(t+1-\alpha)}.
\label{FrFac}
\end{equation}
Taking into account that 
falling factorial-law memory is asymptotically power-law memory (see
Fig.~\ref{FFtoP} and Eq.~(\ref{GammaLimit}) in this paper), we may expect that 
fractional difference maps have properties similar to the  properties of 
fractional maps.  Differences in the maps' properties 
due to the differences in the 
weights of the recent (with $(n-j)/n <<1$)  states (a state is a set 
of variables which defines a system) at the time  instants $t_j$
in the definition of the present state at time $t_n$ 
should be significant when $\alpha \in (0,1)$ 
(especially when $\alpha \rightarrow +0$), as it can be seen from 
Fig.~\ref{KcFig} and comparison of Figs.~\ref{FFtoP}~a~and~b.

The goal of the present paper is to conduct an investigation of  fractional
difference maps consistent with the previous research of
fractional maps
\cite{MyReview,TZ,FM1,FM2,FM3,FM4,FM5,FM6,MEDNC,MEChaos}
and make a step towards the understanding of the general properties of 
systems with asymptotically power-law memory. This will also lead 
to the understanding of the general properties of solutions of  
nonlinear fractional difference equations.
In our investigation we use  the
fractional difference Caputo Standard 
$\alpha$-Family of Maps ($\alpha$FM) introduced in \cite{IFAC}, 
which is an extension of the
regular Standard Map \cite{Chirikov,LL,ZasBook}.
A paper on the  fractional
difference Caputo Logistic $\alpha$FM introduced in \cite{IFAC}, 
which is an extension of the
regular Logistic Map \cite{May},
will be the subject of a separate publication. 

In the next section (Sec.~\ref{FFDFM}) we will recall the notions of 
fractional and fractional difference Caputo  $\alpha$FMs and in
the following Sec.~\ref{Num} we'll compare properties of the   
fractional and fractional difference Caputo Standard $\alpha$FMs.

\section{Fractional and Fractional Difference  
Caputo Standard $\alpha$-Families of Maps}
\label{FFDFM}

\subsection{Fractional Caputo Standard 
$\alpha$-Family of Maps}
\label{FFM}

Fractional $\alpha$FMs were introduced and investigated in \cite{MEDNC,MEChaos} 
(see also review \cite{MyReview}). They are identical
to the following equation:  
\begin{equation}
\frac{d^{\alpha}x}{dt^{\alpha}}+G_K(x(t- \Delta)) \sum^{\infty}_{k=-\infty} \delta \Bigl(t-(k+\varepsilon)
\Bigr)=0,   
\label{UM1D2Ddif}
\end{equation}                                                       
where $\varepsilon > \Delta > 0$,  $\alpha \in \mathbb{R}$, $\alpha>0$, 
$\varepsilon  \rightarrow 0$, with the initial conditions corresponding to
the type of a fractional derivative to be used. $G_K(x)$ is a nonlinear
function which depends on the nonlinearity parameter $K$. 

The fractional Caputo Standard  $\alpha$FM is generated by 
\begin{itemize}
\item
{
using in Eq.~(\ref{UM1D2Ddif}) the left-sided Caputo fractional derivative 
\cite{SKM,KST,Podlubny} 
{\setlength\arraycolsep{0.5pt}
\begin{eqnarray}
&&_0^CD^{\alpha}_t x(t)=_0I^{N-\alpha}_t \ D^N_t x(t) 
\label{Cap} \\  
&&=\frac{1}{\Gamma(N-\alpha)}  \int^{t}_0 
\frac{ D^N_{\tau}x(\tau) d \tau}{(t-\tau)^{\alpha-N+1}},  \quad ( N=\lceil \alpha \rceil),
\nonumber
\end{eqnarray}
}
where $N \in \mathbb{Z}$,  
$D^N_t=d^N/dt^N$, $ _0I^{\alpha}_t$ is a fractional integral,
$\Gamma()$ is the gamma function;
}
\item
{
using the initial conditions 
\begin{equation}
(D^{k}_tx)(0+)=b_k, \ \ \ k=0,...,N-1;
\label{UM1D2DdifIC}
\end{equation} 
}
\item
{
and assuming 
\begin{equation}
G_K(x)=K \sin(x).
\label{SFM}
\end{equation} 
}
\end{itemize}
Then, after the introduction  $x^{(s)}(t)=D^s_tx(t)$, 
integration of  Eq.~(\ref{UM1D2Ddif}) produces 
{\setlength\arraycolsep{0.5pt}
\begin{eqnarray}
&&x^{(s)}_{n+1}= \sum^{N-s-1}_{k=0}\frac{x_0^{(k+s)}}{k!}(n+1)^{k} \nonumber \\ 
&&-\frac{K}{\Gamma(\alpha-s)}\sum^{n}_{k=0} \sin(x_k) (n-k+1)^{\alpha-s-1},
\label{FrCSMMapx}
\end{eqnarray} 
}
where $s=0,1,...,N-1$.
We call the map Eq.~(\ref{FrCSMMapx})  
the fractional Caputo Standard  $\alpha$FM because
in the 2D case ($\alpha=2$) it can be reduced to the regular Standard Map
(see \cite{Chirikov}),
which on a torus can be written as
\begin{equation}
p_{n+1}= p_{n} - K \sin(x_n), \ \ \ ({\rm mod} \ 2\pi ), 
\label{SMp}
\end{equation}
\begin{equation}
x_{n+1}= x_{n}+ p_{n+1}, \ \ \ ({\rm mod} \ 2\pi ).
\label{SMx}
\end{equation}

In \cite{MyReview,MEDNC,MEChaos} the Caputo Standard $\alpha$FM 
was investigated in detail for the case $\alpha \in [0,2]$ that is
important in applications.
\begin{itemize}
\item{
For $\alpha=0$ the Caputo Standard  $\alpha$FM is 
identically zero: $x_n=0$.
}
\item{
For $0<\alpha <1$ the Caputo Standard $\alpha$FM  is           
\begin{equation}
x_{n}=  x_0- 
\frac{K}{\Gamma(\alpha)}\sum^{n-1}_{k=0} \frac{\sin{(x_k)}}{(n-k)^{1-\alpha}},
 \   \  ({\rm mod} \ 2\pi ).
\label{FrCMapSM}
\end{equation}
}
\item{
For $\alpha=1$  the 1D Standard Map 
is the Circle Map with zero driving phase
\begin{equation}
x_{n+1}= x_n - K \sin (x_n), \ \ \ \ ({\rm mod} \ 2\pi ). 
\label{SM1D} 
\end{equation}
}
\item{
For $1<\alpha <2$ the Caputo Standard $\alpha$FM is
{\setlength\arraycolsep{0.5pt}  
\begin{eqnarray}
&& p_{n+1} = p_n - 
 \frac{K}{\Gamma (\alpha -1 )} 
\Bigl[ \sum_{i=0}^{n-1} V^2_{\alpha}(n-i+1) \sin (x_i)  \nonumber \\
&& + \sin (x_n) \Bigr],\ \ ({\rm mod} \ 2\pi ),  \label{FSMCp}  \\ 
&& x_{n+1}=x_n+p_0-
\frac{K}{\Gamma (\alpha)} 
\sum_{i=0}^{n} V^1_{\alpha}(n-i+1) \sin (x_i), \nonumber \\ 
&& ({\rm mod} \ 2\pi ), 
\label{FSMCx}
\end{eqnarray}
}
where $V^k_{\alpha}(m)=m^{\alpha -k}-(m-1)^{\alpha -k}$.
}
%\end{itemize} 
\item{
For $\alpha=2$  the Caputo Standard Map is the regular Standard Map as in
Eqs.~(\ref{SMp})~and~(\ref{SMx}) above.
} 
\end{itemize} 

\subsection{Fractional Difference  
Caputo Universal $\alpha$-Family of Maps}
\label{FDO}

As we mentioned in the Introduction, fractional difference calculus is a
subject of extensive current research. To introduce the  
fractional difference  Caputo Standard $\alpha$-Family of Maps
we will use only one theorem (Theorem 3 from \cite{IFAC}):
\begin{theorem}
 For $\alpha \in \mathbb{R}$, $\alpha \ge 0$ the Caputo-like 
difference equation 
\begin{equation}
_0^C\Delta^{\alpha}_t x(t) = -G_K(x(t+\alpha-1)),
\label{LemmaDif_n}
\end{equation}
where $t\in \mathbb{N}_{m}$, with the initial conditions 
 \begin{equation}
\Delta^{k} x(0) = c_k, \ \ \ k=0, 1, ..., m-1, \ \ \ 
m=\lceil \alpha \rceil
\label{LemmaDifICn}
\end{equation}
is equivalent to the map with falling factorial-law memory
{\setlength\arraycolsep{0.5pt}   
\begin{eqnarray} 
&&x_{n+1} =   \sum^{m-1}_{k=0}\frac{\Delta^{k}x(0)}{k!}(n+1)^{(k)} 
\nonumber \\
&&-\frac{1}{\Gamma(\alpha)}  
\sum^{n+1-m}_{s=0}(n-s-m+\alpha)^{(\alpha-1)} 
G_K(x_{s+m-1}), 
\label{FalFacMap}
\end{eqnarray}
}
where $x_k=x(k)$, 
which we will call the  fractional difference Caputo  Universal  
$\alpha$-Family of Maps.
\end{theorem}
In this theorem $_0^C\Delta^{\alpha}_t$ is defined by Anastassiou  
\cite{Anastas} for noninteger $\alpha>0$ fractional (left) 
Caputo difference operator as
{\setlength\arraycolsep{0.5pt}   
\begin{eqnarray} 
&&_a^C\Delta^{\alpha}_t x(t) =  _a\Delta^{-(m-\alpha)}_{t}\Delta^{m} x(t)\nonumber \\
&& =\frac{1}{\Gamma(m-\alpha)} \sum^{t-(m-\alpha)}_{s=a}(t-s-1)^{(m-\alpha-1)} 
\Delta^m x(s),
\label{FDC}
\end{eqnarray}
} 
where $\Delta^{m}$ is the $m$-th power of the forward difference operator
defined as $\Delta x(t)=x(t+1)-x(t)$, extended in \cite{IFAC} to all real 
$\alpha \ge 0$ by defining $_a^C\Delta^{m}_t x(t) = \Delta^m x(t)$ 
for $m \in \mathbb{N}_0$, where $\mathbb{N}_t=\{t,t+1, t+2, ...\}$.

The family of maps  Eq.~(\ref{FalFacMap}) is called universal because 
in the 2D case ($\alpha=2$) after the introduction $p_n=\Delta x_{n-1}$ 
and with the assumption $G_K(x)=KG(x)$ it can be written as the regular
Universal Map (see, e.g. \cite{ZasBook})
\begin{equation}
p_{n+1}= p_{n} - K G(x_n),  
\label{UMp}
\end{equation}
\begin{equation}
x_{n+1}= x_{n}+ p_{n+1}.
\label{UMx}
\end{equation}

\subsubsection{Integer-Dimensional Difference Universal Maps}
\label{UniversalInteger}

In the case of the integer $\alpha=m$ Eq.~(\ref{LemmaDif_n}) can be written as 
\begin{equation}
\Delta^{m} x_n = -G_K(x_{n+m-1}),
\label{ILemmaDif_n}
\end{equation}
which for $m=0$ assumes the form
\begin{equation}
x_{n+1} = -G_K(x_{n})
\label{m0}
\end{equation}
and for $m=1$ assumes the form
\begin{equation}
x_{n+1} = x_n-G_K(x_{n}).
\label{m1}
\end{equation}
For $m>1$ let's define 
\begin{equation}
x_n^0=x_n,  \ \ x_n^s=\Delta x_{n-1}^{s-1}, \ \ s=1,2, ..., m-1.
\label{xs}
\end{equation}
Then $ x_n^s=\Delta^s x_{n-s}^0$ and
Eq.~(\ref{ILemmaDif_n}) is equivalent to the $m$-dimensional 
map 
\begin{equation}
x_{n+1}^s=\sum^{m-1}_{k=s}x_n^k  -G_K(x_n^0),  \ \ s=0,1, ..., m-1,
\label{mDim}
\end{equation}
which Jacobian $m \times m$ matrix 
$J_{(x_{n+1}^0,x_{n+1}^1,...,x_{n+1}^{m-1})} (x_{n}^0,x_{n}^1,...,x_{n}^{m-1})$ is
\[ \left| \begin{array}{ccccccccc}
1-\dot{G}_K(x_n^0) & 1 & 1  & ... & 1 &
... & 1 & 1 \\
-\dot{G}_K(x_n^0) & 1 & 1  & ... & 1 &
... & 1 & 1 \\
-\dot{G}_K(x_n^0) & 0 & 1  & ... & 1 &
... & 1 & 1 \\
... & ... & ... & ... & ... &
... & ... & ... \\
-\dot{G}_K(x_n^0) & 0 & 0 & ... & 0 &
... & 0 & 1 \end{array} \right| . \]
The first column of this matrix can be written as the sum of 
the column with one in the
first row and the remaining zeros and the column which is equal to
$-\dot{G}_K(x)$ times the last column. Determinants of the corresponding
matrices are 1 and 0; this is why the Jacobian determinant is equal 
to one and the map, similarly to the m-dimensional Universal
Map (Eqs.~(13) and (14) in \cite{MEChaos}), 
is the m-dimensional volume preserving map.
The m-dimensional difference Universal and 
Universal Maps are identical only for the cases $m=1$ and $m=2$.

\subsection{Fractional Difference  
Caputo Standard $\alpha$-Family of Maps}
\label{FD0s}

For $G(x)=\sin(x)$ the map Eqs.~(\ref{UMp})~and~(\ref{UMx})   
is equivalent to the regular Standard Map
Eqs.~(\ref{SMp})~and~(\ref{SMx}). This is why we will call the map
Eq.~(\ref{FalFacMap})  with $G_K(x)=K\sin(x)$ 
{\setlength\arraycolsep{0.5pt}   
\begin{eqnarray} 
&&x_{n+1} =   \sum^{m-1}_{k=0}\frac{\Delta^{k}x(0)}{k!}(n+1)^{(k)} 
\nonumber \\
&&-\frac{K}{\Gamma(\alpha)}  
\sum^{n+1-m}_{s=0}(n-s-m+\alpha)^{(\alpha-1)} 
\sin(x_{s+m-1}) 
\label{FalFacSMap}
\end{eqnarray}
}
the fractional difference Caputo Standard $\alpha$FM.
\begin{itemize}
\item{
In the case $\alpha=0$ the 0D Standard Map turns into the Sine Map (see,
e.g., \cite{Sin})
\begin{equation}
x_{n+1} = -K\sin(x_n), \ \ \ ({\rm mod} \ 2\pi ).
\label{SM0D}
\end{equation} 
}
\item{
For $0<\alpha<1$ the fractional difference Caputo Standard  $\alpha$FM
is 
{\setlength\arraycolsep{0.5pt}   
\begin{eqnarray} 
&&x_{n+1} =  x_0  
\label{SMlt1} \\
&& -\frac{K}{\Gamma(\alpha)}
\sum^{n}_{s=0}\frac{\Gamma(n-s+\alpha)}{\Gamma(n-s+1)}\sin (x_s) 
, \ \ \ ({\rm mod} \ 2\pi ),
\nonumber
\end{eqnarray}
}
which after the $\pi$-shift of the independent variable 
$x\rightarrow x+\pi$ coincides with the ``fractional sine map''
proposed in \cite{Fall}.
}
\item{$\alpha=1$ difference  Caputo 
Standard   $\alpha$FM is identical to the Circle Map with zero driven phase
Eq.~(\ref{SM1D}). The map considered in \cite{Fall} 
\begin{equation}
x_{n+1}= x_n + K \sin (x_n), \ \ \ \ ({\rm mod} \ 2\pi )
\label{SM1DNeg} 
\end{equation}
is obtained from this map by the substitution $x \rightarrow x+\pi$.
}
\item{
For $1<\alpha<2$ the fractional difference Caputo Standard  $\alpha$FM
is 
{\setlength\arraycolsep{0.5pt}   
\begin{eqnarray} 
&&x_{n+1} =  x_0 +\Delta x_0 (n+1) -\frac{K}{\Gamma(\alpha)}
\label{SMgt1} \\
&& 
\times \sum^{n-1}_{s=0}\frac{\Gamma(n-s+\alpha-1)}{\Gamma(n-s)}\sin (x_{s+1}) 
, \ \ ({\rm mod} \ 2\pi ),
\nonumber
\end{eqnarray}
}
which after the introduction of  $p_n=\Delta x_{n-1}$ can be written 
as a 2D map with memory
{\setlength\arraycolsep{0.5pt}   
\begin{eqnarray} 
&&p_{n} =  p_1 -\frac{K}{\Gamma(\alpha-1)}
\label{SMgt1p} \\
&&
\times \sum^{n}_{s=2}\frac{\Gamma(n-s+\alpha-1)}
{\Gamma(n-s+1)}\sin (x_{s-1}) 
, \ \ ({\rm mod} \ 2\pi ),
\nonumber  \\
&& x_n=x_{n-1}+p_n, \ \ ({\rm mod} \ 2\pi ),  \ \ n \ge 1,
\label{SMgt1x}
\end{eqnarray}
}
which in the case $x_0=0$ is identical to the 
"fractional standard map" introduced in \cite{Fall} (Eq.~(18) 
with $\nu=\alpha-1$ there). 
}
\item{The $\alpha=2$  difference  Caputo 
Standard $\alpha$FM is  
the regular Standard Map Eqs.~(\ref{SMp})~and~(\ref{SMx}).
}
\end{itemize}

\section{Properties of the Fractional 
and Fractional Difference 
Caputo Standard $\alpha$FM}
\label{Num}

\begin{figure}[!t]
%\centering
%\includegraphics[width=3in]{Fig1.eps}
\includegraphics{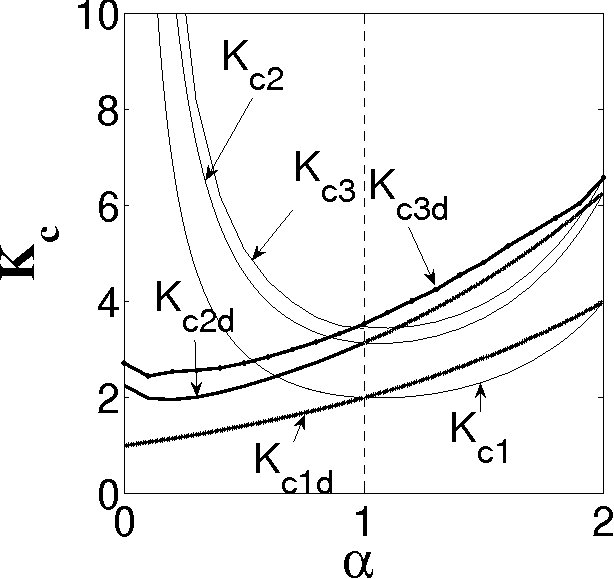}
\vspace{-0.25cm}
\caption{$\alpha-K$ (bifurcation) diagrams  for the Caputo 
(thin lines) and  Fractional  Difference Caputo (bold lines and extra index
``d'') Standard $\alpha$FMs.
The $(0,0)$ fixed point is stable in the area below the curve $K_{c1}$  
($K_{c1d}$ for the difference map). 
The period two ($T=2$) symmetric   sink ($x_{n+1}=-x_{n}$) is stable in 
the area between $K_{c1}$ and  $K_{c2}$ ($K_{c1d}$ and  $K_{c2d}$ for the
difference map). $K_{c3}$  ($K_{c3d}$ for the difference map) is the
border with chaos (above this curve). Cascade of bifurcations type
trajectories can be found in the area near this curve (below it).
  }
\label{KcFig}
\end{figure}

The main properties of the  Fractional Difference Caputo Standard
$\alpha$FM and their differences  from the properties of the
 Fractional  Caputo Standard   $\alpha$FM for $\alpha \in (0,2)$ are summarized in 
$\alpha-K_c$ diagram Fig.~\ref{KcFig}.

\subsection{Integer $\alpha$}

\subsubsection{The Sine Map ($\alpha=0$) }

\begin{figure}[!t]
%\centering
%\includegraphics[width=7in]{Fig2.eps}
\includegraphics{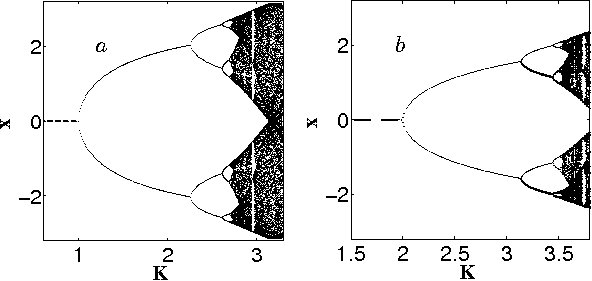}
\vspace{-0.25cm}
\caption{Bifurcation diagrams for a). The Sine Map (difference map with 
$\alpha=0$) Eq.~(\ref{SM0D}) and
b). The Circle Map ($\alpha=1$) with zero driven phase Eq.~(\ref{SM1D}).
  }
\label{Bif0and1}
\end{figure}

The bifurcation diagram for the case $\alpha=0$, the Sine Map Eq.~(\ref{SM0D}), 
with $|K| \le 2\pi$ can be found in \cite{Sin} and with $K \in [0.6,3.3]$
in Fig.~\ref{Bif0and1}a.
%\begin{figure*}[!t]
%\centering
%\includegraphics[width=6in]{Fig2.eps}
%\caption{Bifurcation diagrams for a). The Sine Map (difference map with 
%$\alpha=0$) Eq.~(\ref{SM0D}) and
%b). The Circle Map ($\alpha=1$) with zero driven phase Eq.~(\ref{SM1D}).
%  }
%\label{Bif0and1}
%\end{figure*}
It is easy to show by means of the standard stability analysis
that the fixed point $x=0$ is a sink for $|K|<1$ and the period two ($T=2$)
point $x_{n+1}=-x_n$ is a sink for $1<K<2.262$ (at $K=2.262$ we have 
$\tan2.029=-2.029$ and $|x_n|=2.029$). At $K=2.262$  two new 
$T=2$ sinks appear, which  later (for
larger $K$) bifurcate and give birth to the $T=4$ sink and so on.
This period doubling cascade of bifurcations process leads 
to the onset of chaos at $K \approx 2.72$.
In  Fig.~\ref{KcFig} the curves $K_{c1d}$, $K_{c2d}$, and $K_{c3d}$ 
intersect the line $\alpha = 0$ at the points 1, 2.262, and 2.72
correspondingly.

\subsubsection{The Circle Map with Zero Driven Phase ($\alpha=1$)}

The  Circle Map  with zero driven phase Eq.~(\ref{SM1D}), which can also
be called the 1D Standard Map, is investigated in \cite{MEDNC,MEChaos}
and for $1.5<K<3.8$ is presented in Fig.~\ref{Bif0and1}b.
In  Fig.~\ref{KcFig} the intersections of  the curves  $K_{c1d}$, $K_{c2d}$, 
and $K_{c3d}$ with the line $\alpha = 1$ take place  at the points 2, 
$\pi$, and 3.532 correspondingly (the same is true for the curves
$K_{c1}$, $K_{c2}$, and $K_{c3}$). 
Here we have to notice that the 
transition at $K=\pi$ is not from a $T=2$ sink to a $T=4$ sink,
but from the  $x_{n+1}=-x_n$ period two sink to two
$x_{n+1}=x_n+\pi$ period two sinks and in order to outline the whole
bifurcation curve one should run computer codes with initial conditions
$\pm x_0$ (something that the authors of \cite{Fall} failed to notice). 
In  Fig.~\ref{Bif0and1}b (and in Fig.~\ref{BifLT1})
two sets of initial conditions correspond to two sets of points:
the regular points ($x_0=0.1$) and the bold points ($x_0=-0.1$). 

\subsubsection{The Standard Map ($\alpha=2$) }

The Standard Map (Chirikov Map) is one of the best-investigated maps
(see \cite{Chirikov,LL}). It demonstrates a universal generic behavior 
of the area-preserving maps whose
phase space is divided into elliptic islands of stability and areas of 
chaotic motion. 
The $(0,0)$ elliptic point becomes unstable
(elliptic-hyperbolic point transition) at $K=4$ and gives birth
to two  elliptic islands around the stable (for $4 < K <2 \pi$)  
$T=2$ antisymmetric ($p_{n+1}=-p_n$, $x_{n+1}=-x_n$) trajectory.
At $ K =2 \pi$ the antisymmetric $T=2$ point turns into two stable $T=2$
points with  $p_{n+1}=-p_n$, $|x_{n+1}-x_n|=\pi$. The following
period doubling cascade of bifurcations  leads to the disappearance of
the islands of stability in the chaotic sea at $K \approx 6.6344$.
In  Fig.~\ref{KcFig} the intersections of  the curves $K_{c1d}$, $K_{c2d}$, 
and $K_{c3d}$ with the line $\alpha = 2$ take place  at the points 4, 
$2\pi$, and 6.6344 correspondingly (the same is true for the curves
$K_{c1}$, $K_{c2}$, and $K_{c3}$).

\subsection{ $0 < \alpha < 1$}
\label{SecLT1}

\begin{figure}[!t]
\centering
\includegraphics[width=5in]{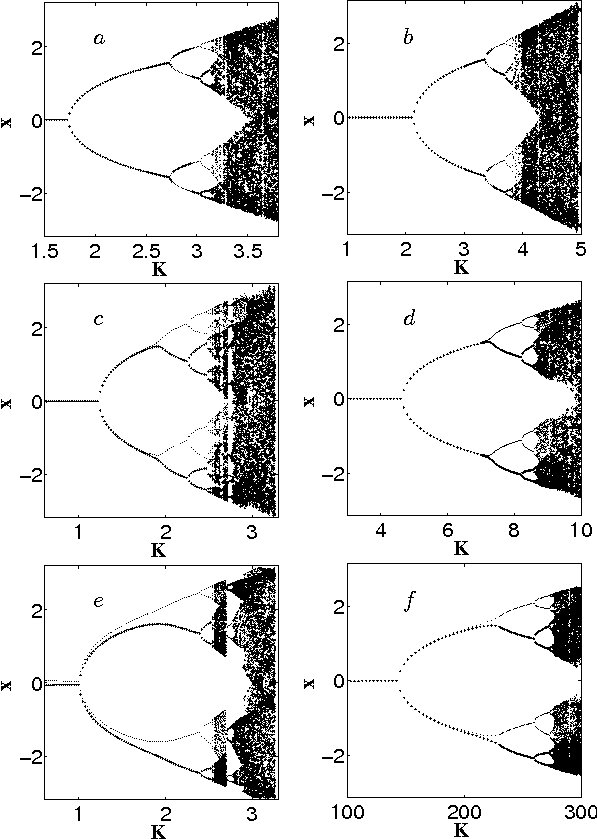}
\vspace{-0.25cm}
\caption{Bifurcation diagrams for the fractional  difference Caputo Standard
$\alpha$FM Eq.~(\ref{SMlt1})  ($a$, $c$, and $e$) and the  fractional  Caputo
Standard $\alpha$FM  Eq.~(\ref{FrCMapSM}) ($b$, $d$, and $f$). The
diagrams were obtained after 5000 iterations with the initial
condition $x_0=0.1$ (regular points) and $x_0=-0.1$ (bold points). 
$\alpha = 0.8$ in $a$ and $b$;  $\alpha = 0.3$ in $c$ and $d$;
$\alpha = 0.01$ in $e$ and $f$.  
}
\label{BifLT1}
\end{figure}

Sample bifurcation diagrams for the fractional  and fractional  difference
Caputo Standard $\alpha$FM with $0 < \alpha < 1$ are presented
in Fig.~\ref{BifLT1}.
One obvious difference between two  $\alpha$FMs is that as $\alpha$
decreases towards zero, bifurcation diagrams of the fractional difference 
maps Figs.~\ref{BifLT1} a, c, and e  contract along the $K$-axis approaching 
the bifurcation diagram of the Sine Map Fig.~\ref{Bif0and1}a, while
the bifurcation diagrams of the fractional maps Figs.~\ref{BifLT1} b, d, and f 
expand along the  $K$-axis.

The complete analysis of these bifurcation diagrams is not a subject 
of the present paper, but we'll outline some analytic results
which were confirmed by the direct simulations of fractional
maps.

\begin{figure}[!t]
%\centering
%\includegraphics[width=3in]{Fig4.eps}
\includegraphics{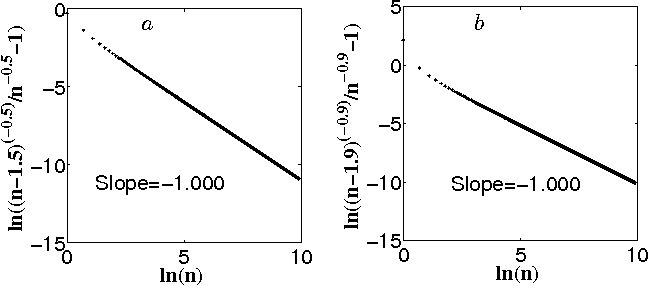}
\caption{Falling factorial $(n+\alpha-2)^{(\alpha-1)}$ to power law 
$n^{\alpha-1}$ ratio.   $\alpha=0.5$ in (a) and  $\alpha=0.1$
in (b). 
}
\label{FFtoP}
\end{figure}

Both maps, Eq.~(\ref{FrCMapSM}) and Eq.~(\ref{SMlt1}), can be written
in the form
\begin{equation}
x_{n}=  x_0- 
\frac{K}{\Gamma(\alpha)}\sum^{n-1}_{k=0} W_{\alpha}(n-k) \sin{(x_k)},
\label{1_2DSinMaps}
\end{equation}
where $W_{\alpha}(s) = s^{\alpha-1}$ for the fractional map and
 $W_{\alpha}(s) ={\Gamma(s+\alpha-1)}/{\Gamma(s)}$ for the fractional
 difference map. Asymptotically, both expressions coincide (see Fig.~\ref{FFtoP}) because
\begin{equation}
\lim_{s \rightarrow
  \infty}\frac{\Gamma(s+\alpha)}{\Gamma(s+1)s^{\alpha-1}}=1,  
\ \ \ \alpha \in  \mathbb{R}.
\label{GammaLimit}
\end{equation}

In the Sine Map and in the Circle Map with zero driven phase
at the point where the $x=0$ sink becomes unstable it gives birth to
a symmetric $T=2$  point in which $x_{n+1}=-x_{n}$.
Following the results of our numeric simulations, let's assume that 
this property persists (asymptotically) for $\alpha \in (0,1)$.
Eq.~(\ref{1_2DSinMaps}) can be written as
{\setlength\arraycolsep{0.5pt}   
\begin{eqnarray} 
&&x_{n+1}=  x_n- 
\frac{K}{\Gamma(\alpha)} \Bigl\{  W_{\alpha}(1) \sin{(x_n)} \nonumber \\
&&+\sum^{n-1}_{k=0} \sin{(x_k)} [W_{\alpha}(n-k+1)-W_{\alpha}(n-k)]\Bigr\}. 
\label{1_2DSinMapsD}
\end{eqnarray}
}
Taking into account that $W_{\alpha}(n-k+1)-W_{\alpha}(n-k) \rightarrow 0$
as $n \rightarrow \infty$, after substitution $j=n-k$ for large $n$ we may write
{\setlength\arraycolsep{0.5pt}   
\begin{eqnarray} 
&&x_n=\frac{K}{2\Gamma(\alpha)}\Bigl\{
  W_{\alpha}(1) \nonumber \\
&&+\sum^{\infty}_{j=1} (-1)^j[W_{\alpha}(j+1)-W_{\alpha}(j)]\Bigr\}\sin{(x_n)},
\label{1_2DSinMapsD1} 
\end{eqnarray}
}
where the alternating series on the right side converges because
its terms converge to 0 monotonically.
This equation has real non-trivial solutions when 
{\setlength\arraycolsep{0.5pt} 
\begin{eqnarray} 
&&K>K_{cr1} \nonumber \\
&&=\frac{2\Gamma(\alpha)}{W_{\alpha}(1)
+\sum^{\infty}_{j=1} (-1)^j[W_{\alpha}(j+1)-W_{\alpha}(j)]}.
\label{1_2DSinMapsKcr}
\end{eqnarray}
}
Numeric calculations of Eq.~(\ref{1_2DSinMapsKcr}) with the
corresponding functions $W_{\alpha}$
were performed to obtain the curves 
$K_{c1}$ and $K_{c1d}$ for $\alpha \in (0,1)$ 
in Fig.~\ref{KcFig} and they were also confirmed by
the  direct numeric simulations of the maps.

The direct numeric simulations of the maps show that for the   
fractional (this is not true for the fractional difference)  
Caputo Standard $\alpha$FM, at the value of $K$ when the antisymmetric 
$T=2$ point becomes unstable, two new $T=2$ sinks appear with the
property $|x_{n+1}-x_n|=\pi$. Then, an asymptotic consideration 
similar to the one presented above leads to 
{\setlength\arraycolsep{0.5pt}   
\begin{eqnarray} 
&&\pm \pi=\frac{K}{\Gamma(\alpha)}\Bigl\{
  W_{\alpha}(1) \nonumber \\
&&+\sum^{\infty}_{j=1} (-1)^j[W_{\alpha}(j+1)-W_{\alpha}(j)]\Bigr\}\sin{(x_n)}
\label{1_2DSinMapsD2} 
\end{eqnarray}
}
and 
\begin{equation} 
K>K_{c2}=\pi K_{c1}/2.
\label{Kcr2}
\end{equation}
The last equation was used to calculate the curve 
$K_{c2}$ in Fig.~\ref{KcFig}. The curves $K_{c2d}$,  $K_{c3}$, and $K_{c3d}$
were obtained by the direct numeric simulations of the maps.

Periodic sinks $x=x_l$ (except the $x=0$ fixed point) exist only 
in the asymptotic sense. Trajectories starting at   $x_l$ jump out of the
sink and then converge asymptotically according to a power law $x-x_l \sim
n^{-\alpha}$.   
\begin{figure}[!t]
%\centering
%\includegraphics[width=3.in]{Fig5.eps}
\includegraphics{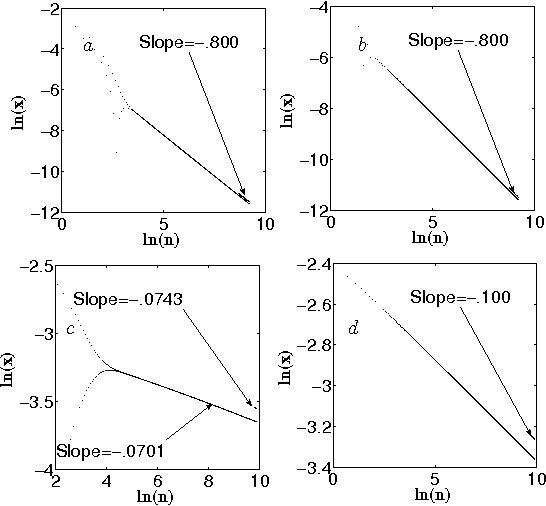}
\caption{Convergence of trajectories to the $x=0$ sink for the 
fractional ($b$ and $d$) and fractional difference ($a$ and $c$) 
Caputo Standard $\alpha$FM ($\alpha \in (0,1)$). In all cases $x_0=0.1$. 
10000 iterations, $\alpha = 0.8$, and $K=1.5$ in $a$  and $b$.
20000 iterations, $\alpha = 0.1$, and $K=1.0$ in $c$  and $d$.  
}
\label{ConvergenceLT1}
\end{figure}
This law of convergence to the $x=0$ sink is demonstrated 
in Fig.~\ref{ConvergenceLT1}. For small $\alpha$ the rate of convergence
is very slow. For the difference map even the rate of convergence
itself is converging to its asymptotic value very slowly 
(Fig.~\ref{ConvergenceLT1}c).  Significance of the 
slow rate convergence  for the explanation of the  fact that in the 
fractional difference Caputo Standard $\alpha$FM with small values of $\alpha$
the bifurcation diagrams depend on the initial conditions 
(Fig.~\ref{BifDifAlpDep}) is not investigated in the present paper.
\begin{figure}[!t]
\centering
\includegraphics[width=5.in]{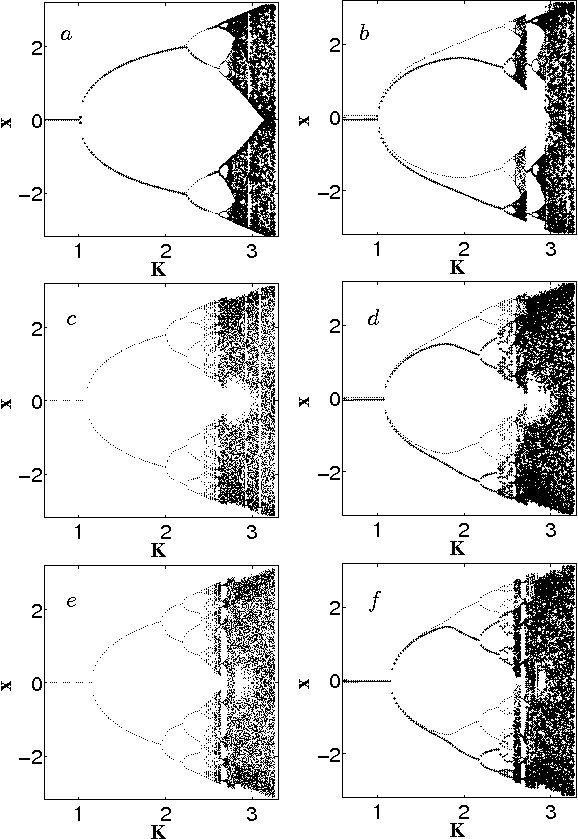}
\caption{Dependence of the  fractional difference Caputo Standard
  $\alpha$FM's bifurcation diagrams on the initial conditions for 
small $\alpha$. $\alpha=10^{-10}$ in $a$ and $b$; $\alpha=.1$ in $c$ and
$d$;  $\alpha=.2$ in $e$ and $f$. 
In $a$ and $b$ the bifurcation diagrams obtained after 200 iterations
for each $K$. In $c$, $d$, $e$, and $f$ the bifurcation diagrams 
obtained after 5000 iterations for each $K$. 
The initial conditions: $x_0=\pm 0.001$ in $a$;  $x_0=\pm 0.00001$ in $c$
and $e$;  $x_0=\pm 0.1$ in $b$, $d$, and  $f$.
}
\label{BifDifAlpDep}
\end{figure}
As can be seen from Fig.~\ref{BifDifAlpDep}, the dependence 
of the bifurcation diagrams of the difference maps on the initial 
conditions is significant for  $\alpha<0.2$. 
\begin{figure}[!t]
%\centering
%\includegraphics[width=3in]{Fig7.eps}
\includegraphics{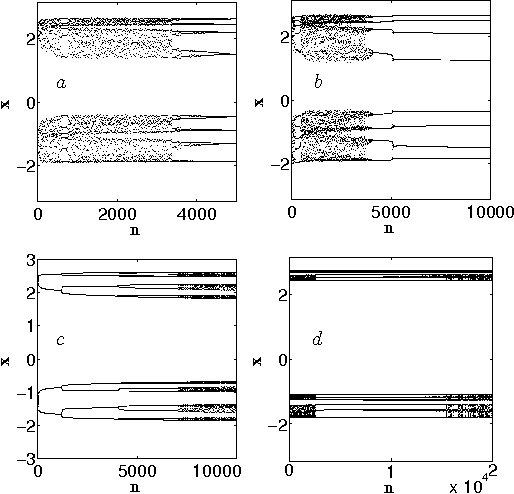}
\caption{Bifurcating trajectories in the  fractional difference Caputo Standard
$\alpha$FM. Each figure represents a single trajectory with:
a). $\alpha=0.2$, $K=2.52$, and $x_0=0.1$; 
b). $\alpha=0.2$, $K=2.55$, and $x_0=0.1$; 
c).   $\alpha=0.1$, $K=2.41$, and $x_0=0.1$; 
d).  $\alpha=0.001$, $K=2.72$, and $x_0=0.003$; 
}
\label{CBTTLT1}
\end{figure}
As in the case of fractional maps \cite{MyReview,MEDNC,MEChaos}, 
individual trajectories 
of the fractional difference Standard  $\alpha$FM with $0 < \alpha <1$ 
in the area of the parameter values for which on the bifurcation diagram
stable periodic $T>2$ sinks exist and the transition to chaos occurs
are cascade of bifurcations type trajectories (CBTT) 
(see Fig.~\ref{CBTTLT1}c). Even more complicated trajectories, including
inverse cascades of bifurcations (Fig.~\ref{CBTTLT1}a) and trajectories with
intermittent chaotic behavior (Fig.~\ref{CBTTLT1}b~and~d), can be found
in the fractional difference Standard  $\alpha$FM.

\begin{figure}[!t]
%\centering
%\includegraphics[width=3in]{Fig8.eps}
\includegraphics{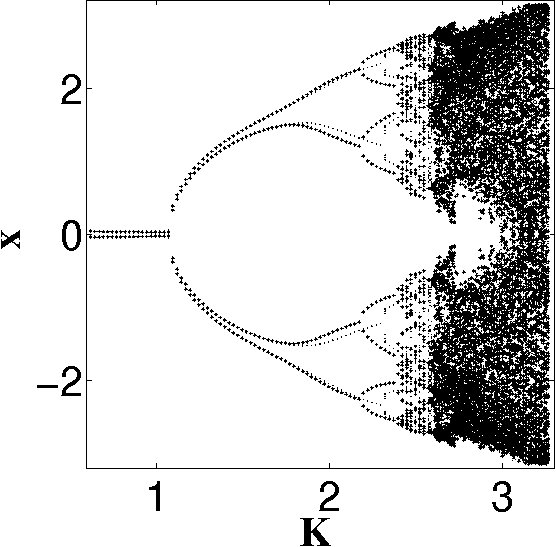}
\caption{Two bifurcation diagrams for the  fractional difference Caputo Standard
$\alpha$FM with   $\alpha=0.1$ and  $x_0=0.1$ calculated after 200
iterations (regular points) and 5000 iterations (bold points).
}
\label{BifLT1shift}
\end{figure}

One of the consequences of the existence of CBTT 
is the dependence of bifurcation
diagrams on the number of iterations after which they are calculated. 
In Fig.~\ref{BifLT1shift} some of the points which after 200 iterations are
$T=2^n$ sinks, after 5000 iterations become  $T=2^{n+1}$ sinks, and
the corresponding bifurcation points shift to the left.   

\subsection{ $1 < \alpha < 2$}

\begin{figure}[!t]
%\centering
%\includegraphics[width=3in]{Fig9.eps}
\includegraphics{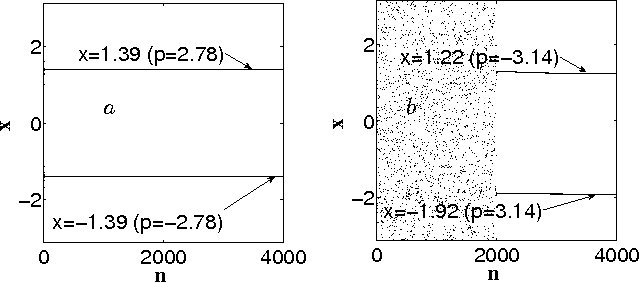}
\caption{Two $T=2$ trajectories for the fractional  difference 
Caputo Standard $\alpha$FM with  $\alpha=1.5$ and  $x_0=0$ and $p_0=0.01$:
a). K=4.0 antisymmetric trajectory $x_{n+1}=-x_n$,  $p_{n+1}=-p_n$; 
b). K=4.74 $\pi$-shift trajectory with  $|x_{n+1}-x_n|=\pi$,  $p_{n+1}=-p_n$. 
}
\label{T2s}
\end{figure}

In this section we'll apply the methods by which the evolution of the 
$(0,0)$ fixed point with the increase in $K$ was investigated for
the fractional Standard Map with  $1 < \alpha < 2$  
in \cite{MyReview,FM1,FM5,FM6,MEDNC,MEChaos} to investigate the 
fractional  difference Caputo Standard $\alpha$FM for  $1 < \alpha < 2$.
As in the fractional Standard Map, when the  $(0,0)$ sink becomes unstable
it gives birth to the  $T=2$ antisymmetric sink  $x_{n+1}=-x_n$,
$p_{n+1}=-p_n$, which later, at $K$ for which
$x_n=\pi/2$, turns into two $\pi$-shift $T=2$ sinks (see Fig~\ref{T2s}).

Assuming the existence of the antisymmetric $T=2$ sink  $x_{n+1}=-x_n$,
$p_{n+1}=-p_n$ and following the same steps as in Sec.~\ref{SecLT1} 
it is easy to derive from Eq.~(\ref{SMgt1p})  for large $n$
{\setlength\arraycolsep{0.5pt}   
\begin{eqnarray} 
&&p_n=\frac{K}{2\Gamma(\alpha-1)}\Bigl\{
  W_{\alpha-1}(1) \nonumber \\
&&+\sum^{\infty}_{j=1} (-1)^j[W_{\alpha-1}(j+1)-W_{\alpha-1}(j)]\Bigr\}\sin{(x_n)},
\label{1_2DSinMapsD1P} 
\end{eqnarray}
}
where, as in Eq~(\ref{1_2DSinMaps}),
$W_{\alpha}(s) ={\Gamma(s+\alpha-1)}/{\Gamma(s)}$.
Eq.~(\ref{SMgt1x})  for large $n$ gives $p_n=2x_n$. 
Then, the equations  defining
the sink $(x_n,p_n)$ are
{\setlength\arraycolsep{0.5pt}   
\begin{eqnarray} 
&&x_n=\frac{K}{4\Gamma(\alpha-1)}\Bigl\{
  W_{\alpha-1}(1) \nonumber \\
&&+\sum^{\infty}_{j=1} (-1)^j[W_{\alpha-1}(j+1)-W_{\alpha-1}(j)]\Bigr\}\sin{(x_n)},
\label{1_2DSinMapsD1Xnnn} \\
&&p_n=2x_n,
\label{1_2DSinMapsD1XPnn}
\end{eqnarray}
}
from which follows that for $1<\alpha<2$
\begin{equation}
K_{c1d}(\alpha)=2K_{c1d}(\alpha-1),
\label{KCR1_2}
\end{equation}
where $K_{c1d}(\alpha-1)$ is defined by  Eq.~(\ref{1_2DSinMapsKcr}). This
result was confirmed by the direct numeric simulations of the maps and used to
calculate the curve $K_{c1d}$ for $\alpha \in (1,2)$ 
in Fig.~\ref{KcFig}.

In a similar way, assuming the existence of the antisymmetric $T=2$ sink  
 $|x_{n+1}-x_n|=\pi$, $p_{n+1}=-p_n$, asymptotically, the equations
defining the sink $(x_n,p_n)$ can be written as 
{\setlength\arraycolsep{0.5pt}   
\begin{eqnarray} 
&&\pm \pi=\frac{K}{2\Gamma(\alpha-1)}\Bigl\{
  W_{\alpha-1}(1) \nonumber \\
&&+\sum^{\infty}_{j=1} (-1)^j[W_{\alpha-1}(j+1)-W_{\alpha-1}(j)]\Bigr\}\sin{(x_n)},
\label{1_2DSinMapsD1Xn} \\
&&p_n=\pm \pi.
\label{1_2DSinMapsShift}
\end{eqnarray}
}
As for the fractional maps, for the fractional difference
maps with  $\alpha \in (1,2)$ the following holds
\begin{equation}
K_{c2d}(\alpha)=\frac{\pi}{2}K_{c1d}(\alpha).
\label{KCR1_2Shift}
\end{equation}
The direct  numeric simulations of the maps confirm this result.
The $K_{c3d}$ curve is obtained by the direct   map's numeric simulations.

\begin{figure}[!t]
%\centering
%\includegraphics[width=3.in]{Fig10.eps}
\includegraphics{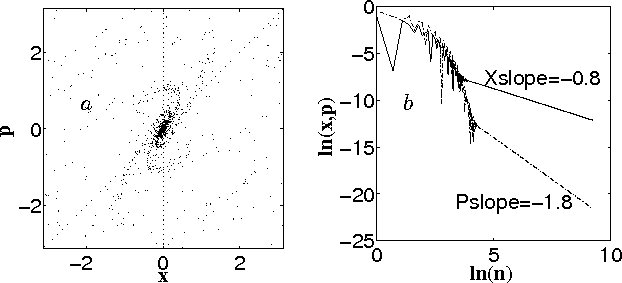}
\caption{The fractional difference Caputo Standard Map 
Eqs.~(\ref{SMgt1p})~and~(\ref{SMgt1x}) with $\alpha=1.8$, 
$K=2.5$: a). Phase space obtained by performing 1000 iterations on each of the 50 
trajectories with $x_0=0$ and $p_1=-3.1415+6.28i/50$, where $0 \le i <50$; 
b). Convergence to the $(0,0)$ sink of a trajectory with $x_0=0$ and
$p_0=0.01$. 
}
\label{Alp1_8K2_5}
\end{figure}

As in the  case of the fractional Caputo Standard Map, the 
trajectories in the  fractional difference Caputo Standard Map
converge to sinks according to a power law. But if in the case of the
fractional Standard Map trajectories converge to the fixed point according to 
$x_n \sim n^{1-\alpha}$ and $p_n \sim n^{1-\alpha}$ 
(see, e.g. Fig.~1e in \cite{FM6}), in the case of the  
fractional difference Standard Map the convergence is according to 
$x_n \sim n^{1-\alpha}$ and $p_n \sim n^{-\alpha}$ (see
Fig.~\ref{Alp1_8K2_5}b). 
As we see, the rate of convergence of the $x$ variable is the same for both
maps. The difference in the  rates of convergence of the $p$ variable
could be  due to the difference in the definitions of momenta $p$ in two
cases. The phase space of the ``fractional Standard Map''
(Eq.~(18) from \cite{Fall}) plotted for the same
$\alpha=1.8$ and $K=2.5$ using 
200 iterations on each of the 400 trajectories with
$(x_0,p_0)=(-3.1415+6.28i/20,-3.1415+6.28j/20)$, where $0 \le i,j <20)$
is identical to the phase space of the fractional difference Standard Map  
Fig.~\ref{Alp1_8K2_5}a; the $\ln(x,p)$ vs. $\ln(n)$ graph for the  
``fractional Standard Map'' with $\alpha=1.8$, $K=2.5$, $x_0=0.3$, and
$p_0=0.1$ is also identical to the one in Fig.~\ref{Alp1_8K2_5}b.
The phase portrait in Fig.~6 from \cite{Fall} for $\alpha=1.8$ and $K=2.5$
with the structure of islands of stability and areas of 
chaotic motion is obviously incorrect.

As in the case of the  fractional Standard Map 
\cite{MyReview,FM1,FM5,FM6,MEDNC,MEChaos}, the most interesting features
of the  fractional difference Standard Map are CBTT and intermittent CBTT
which appear below the border with chaos (curve $K_{c3d}$ in Fig.~\ref{KcFig}). 
\begin{figure}[!t]
\centering
\includegraphics[width=5in]{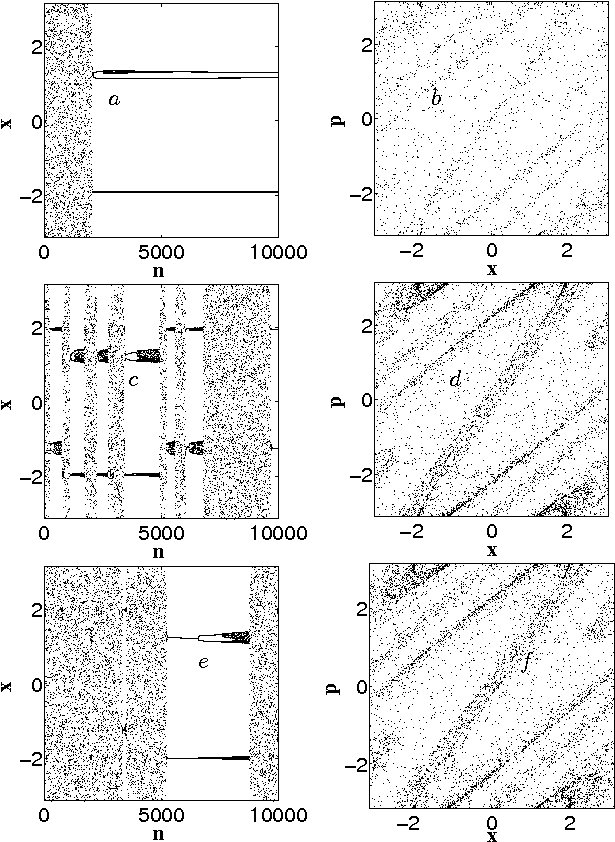}
\caption{Three single trajectories in  the  fractional difference Standard
Map below the border with chaos in phase space ($b$, $d$, and $f$) 
and in $x$ vs. $n$ graphs 
($a$, $c$, and $e$) with the initial conditions $x_0=0$ and $p_0=0.01$. 
$\alpha=1.7$ and $K=5.43$ in $a$ and $b$;  $\alpha=1.5$ and $K=4.82$ in $c$
and $d$;  $\alpha=1.5$ and $K=4.92$ in $e$ and $f$.
}
\label{CBTT1}
\end{figure}

As in the fractional Caputo Standard Map, in the fractional difference 
Caputo Standard Map with $\alpha \in (1,2)$ intermittent CBTT can be found
in $x$ vs. $n$ plots and  reveal themselves best in the middle of the 
$(1,2)$ interval when $\alpha \approx 1.5$
(Fig.~\ref{CBTT1}~$a$,~$c$,~and~$e$). In phase space intermittent CBTT 
are presented as dense dark areas embedded into chaotic attractors near
points where $p=\pm \pi$ (Fig.~\ref{CBTT1}~$b$,~$d$,~and~$f$).
 \begin{figure}[!t]
\centering
\includegraphics[width=5in]{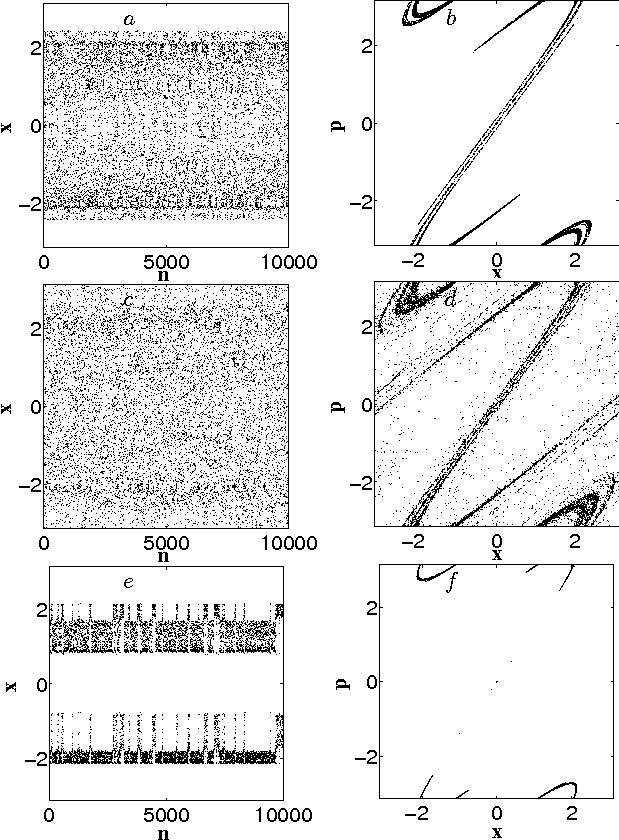}
\caption{Three single trajectories in  the  fractional difference Standard
Map below the border with chaos in phase space ($b$, $d$, and $f$) 
and in $x$ vs. $n$ graphs 
($a$, $c$, and $e$) with the initial conditions $x_0=0$ and $p_0=0.01$. 
$\alpha=1.3$ and $K=4.3$ in $a$ and $b$;  $\alpha=1.3$ and $K=4.45$ in $c$
and $d$;  $\alpha=1.1$ and $K=3.8$ in $e$ and $f$.
}
\label{CBTT2}
\end{figure}
For small $\alpha$ (close to one) as $K$ increases towards the chaotic 
area periodic trajectories  turn into chaotic attractors (Fig.~\ref{CBTT2}).

\section{Conclusion}

The main conclusion based on the results of the presented research 
is that systems with asymptotically power law-memory, similar to systems
with power-law memory, demonstrate behaviors different from the
behaviors of systems with no memory. The new properties include existence of
attracting and intermittent cascade of bifurcations type trajectories,
a common pattern in dependence of bifurcation diagrams on the 
memory parameter $\alpha$, and  non-uniqueness of
solutions (intersection of trajectories and overlapping of attractors)
(see also \cite{MyReview,FM6}). 

The quanitative differences of  properties of  
falling factorial-law memory maps from  power-law memory maps are
the results of the differences in
weights of the recent states in the definition of the present state 
and are significant when $\alpha \rightarrow +0$.
Behavior of systems with small values of $\alpha$ appears to be the 
most interesting (see Figs.~\ref{BifLT1}~e,~f, \ref{BifDifAlpDep}), 
\ref{CBTTLT1}, and \ref{BifLT1shift}. It is interesting to notice that 
the case of small $\alpha$ plays an important role in biological
applications (see, e.g., \cite{MyReview}).
It has been shown recently \cite{L1,L2} that processing of
external stimuli by individual neurons can be described by fractional 
differentiation. The orders of fractional derivatives $\alpha$  
obtained for different types of neurons fall within the interval $[0, 1]$. 
For neocortical pyramidal neurons it is quite small $\alpha \approx 0.15$.
We suggest that it will be important for biological applications to
conduct more theoretical research of the maps with small $\alpha$ and
to make a comparison with experimental biological results.

\section*{Acknowledgments}

The author acknowledges support from the Joseph Alexander Foundation,
Yeshiva University.  
The author expresses his gratitude to E. Hameiri, H. Weitzner,
and  G. Ben Arous
for the opportunity to complete this work at the Courant Institute     
and to V. Donnelly for technical help.

%%%%%%%%%%%%%%%%%%%%%%%%%%%%%%%%%%%%%%%%%%%%%%%%%%%%%%%%%%%%%%%%%%%%%%%%%%%%%%%%%

\end{document}